\documentclass[prl,twocolumn,floatfix,groupedaddress,
nofootinbib,showpacs,preprintnumbers,amsmath,amssymb,amsfonts,superscriptaddress]
{revtex4}
\usepackage{graphicx}
\usepackage{mathrsfs}
\usepackage{bm}

\begin{document}

\title{The Smoothness of Physical Observables}
\author{J. Piekarewicz\footnote{\tt e-mail: jpiekarewicz@fsu.edu}}
\affiliation{Department of Physics, Florida State University,
             Tallahassee, FL {\sl 32306}}
\author{M. Centelles\footnote{\tt e-mail: mariocentelles@ub.edu}}
\affiliation{Departament d'Estructura i Constituents de la Mat\`eria
and Institut de Ci\`encies del Cosmos, Facultat de F\'{\i}sica,
Universitat de Barcelona, Diagonal {\sl 647}, {\sl E-08028} Barcelona,
Spain} 
\author{X. Roca-Maza\footnote{\tt e-mail: roca@ecm.ub.es}}
\affiliation{Departament d'Estructura i Constituents de la Mat\`eria
and Institut de Ci\`encies del Cosmos, Facultat de F\'{\i}sica,
Universitat de Barcelona, Diagonal {\sl 647}, {\sl E-08028} Barcelona,
Spain} 
\author{X. Vi\~nas\footnote{\tt e-mail: xavier@ecm.ub.es}}
\affiliation{Departament d'Estructura i Constituents de la Mat\`eria
and Institut de Ci\`encies del Cosmos, Facultat de F\'{\i}sica,
Universitat de Barcelona, Diagonal {\sl 647}, {\sl E-08028} Barcelona,
Spain} 

 \date{\today}
 \bigskip
 \begin{abstract}
 \medskip
The Garvey-Kelson (GK) relations are powerful algebraic expressions
connecting the masses of neighboring atomic nuclei and derived under
reasonable physical assumptions. In this contribution we show that
these relations are even more general than originally assumed as their
validity is conditioned by the {\sl smoothness} of the underlying
nuclear mass function and nothing else. Based on the assumed
smoothness of the underlying physical function, the main conclusions
from this study are: (1) the GK relations are model independent; (2)
{\sl any} slowly-varying physical observable satisfies the GK
relations; and (3) the accuracy of the GK relations may be
systematically improved.  Examples from nuclear physics (charge radii)
and from particle physics (baryon masses) are used to illustrate the
flexibility of the approach.
 \end{abstract}
\maketitle 

Nuclear observables display a remarkably smooth dependence on atomic
number ($Z$), neutron number ($N$), and baryon number
($A\!=\!Z\!+\!N$). Perhaps the best known example of such a smooth
behavior is the nuclear mass formula of
Weizs\"acker~\cite{Weizsacker:1935} and Bethe and
Bacher~\cite{Bethe:1936}.  In a formulation that dates back to 1935-36
and that has survived the test of time, the atomic nucleus is modelled
as an incompressible liquid drop consisting of two strongly
interacting quantum fluids (one neutral and one charged). The mass
formula includes volume, surface, Coulomb, and asymmetry terms that
vary smoothly with $A$ and $Z$.  When these four empirical
coefficients are fitted to the enormous database of ground-state
masses (of more than 2000 nuclei~\cite{Audi:2002rp}) a rms deviation
of about 3 MeV is obtained.  Most of the 3 MeV discrepancy can be
explained by invoking physics that ceases to be smooth, such as
pairing correlations and shell corrections.  Indeed, highly successful
modern mass formulas that incorporate these effects are able to reduce
the rms deviation by almost an order of magnitude, to about
300~keV~\cite{Moller:1993ed,Duflo:1995}.

An approach that has been recently revitalized, both because of an
interest in understanding any inherent limitation in the nuclear-mass
models as well as because of its astrophysical applications, is the
one by Garvey and Kelson~\cite{Garvey:1966zz,GARVEY:1969zz,
Barea:2005fz,Barea:2008zz,Morales:2009pq}.  The {\sl Garvey-Kelson (GK)
relations} are powerful algebraic expressions connecting masses of
neighboring nuclei. Based on a few simple assumptions about the
underlying nuclear dynamics, Garvey, Kelson, and collaborators derived
the following two relations connecting the masses $M$ of six neighboring
nuclei~\cite{Garvey:1966zz,GARVEY:1969zz}:
\begin{subequations}
 \begin{align}
   \Delta M_{6}^{(1)}(N,Z) 
   &\equiv M(N+2,Z-2)-M(N,Z)
   \nonumber \\       
   & +M(N,Z-1)-M(N+1,Z-2)
   \nonumber \\       
   &+M(N+1,Z)-M(N+2,Z-1) = 0 \,,
  \label{GK1} \\
   \Delta M_{6}^{(2)}(N,Z) 
   &\equiv M(N+2,Z)-M(N,Z-2)
   \nonumber \\       
   & +M(N+1,Z-2)-M(N+2,Z-1)
   \nonumber \\       &+M(N,Z-1)-M(N+1,Z) = 0\,.
  \label{GK2}
 \end{align}
 \label{GKs}
\end{subequations}
It is pertinent to note that in their original publication Garvey and
Kelson made the following germane statement: ``No assumptions are made
about the quantitative aspects of the description except that the
position of the single-particles levels, and the residual interactions
between nucleons in them, {\sl vary slowly} with atomic
number''~\cite{Garvey:1966zz}.  In some sense, this statement is
reminiscent of the role that symmetries and group theory have played
in Nuclear Physics and that had the late Marcos Moshinsky as one of
its leading exponents (see, for example, Ref.~\cite{Moshinsky:1969}).
Rather than attempting an exact solution of the difficult many-body
problem, group-theoretical approaches assume an underlying symmetry of
the (often unknown) Hamiltonian. Based on such an assumed symmetry,
powerful relations can then be derived among the various states of the
Hamiltonian that are connected through the symmetry. It is the aim of
the present Letter to explore the consequences of the {\sl
slowly-varying} dynamics proposed by Garvey and Kelson and to
establish the power and generality of the GK relations, assuming
only the smoothness of the underlying physical observables.

In particular, it will be shown that the linear combinations of masses
chosen by Garvey and Kelson are proportional to the third-order
derivative of the underlying nuclear mass function $M(N,Z)$
~\cite{deShalit:1974}. That is, the selected combinations of masses
are insensitive to the underlying mass function as well as to its
first and second derivatives. It is evident then, that as long as the
mass function is slowly varying, the GK relations will be satisfied to
a very good approximation. For example, if $M(N,Z)$ is represented by
the smooth liquid-drop formula, then the GK mass relations are
satisfied to better than 150 keV for $A\!\ge\!50$ and to better than
50 keV for $A\!\ge\!100$, for nuclei close to the line of stability.
When using the available experimental database of nuclear masses, 
Barea and collaborators~\cite{Barea:2005fz,Barea:2008zz,Morales:2009pq} 
have shown that the GK relations are accurately satisfied throughout the 
periodic table (on average, to about 180--200 keV).
Thus, it was concluded that the GK relations may provide a unique 
tool to test and improve current nuclear mass 
formulas~\cite{Barea:2008zz,Morales:2009pq}---especially in the
as yet experimentally unexplored regions of exotic nuclei. 

To express the GK relations in terms of the third-order derivatives of
the nuclear mass function $M(N,Z)$, both $N$ and $Z$ are treated as
continuous variables so that a Taylor series expansion around the
reference point $(N,Z)$ may be performed. It is then a simple algebraic 
exercise to show that the two GK relations are identically equal to:
\begin{subequations}
 \begin{align}
 \Delta M_{6}^{(1)}(N,Z) &= 
   \frac{\partial^{3}M}{\partial Z^{2}\partial N}  
  -\frac{\partial^{3}M}{\partial Z\partial N^{2}} 
  +\mathcal{O}(\partial^{4}M) \;,
   \label{GK1Approx}\\
 \Delta M_{6}^{(2)}(N,Z) &= 
   \frac{\partial^{3}M}{\partial Z^{2}\partial N}  
  +\frac{\partial^{3}M}{\partial Z\partial N^{2}} 
  +\mathcal{O}(\partial^{4}M) \;.
   \label{GK2Approx}
 \end{align}
 \label{GKsApprox}
\end{subequations}
Whereas Garvey and Kelson invoked reasonable physical arguments to 
construct the two linear combinations of masses given in Eq.~(\ref{GKs}), 
the previous result shows that the GK relations are both more general and 
more powerful than originally assumed. Indeed, the GK relations are model 
independent as they are insensitive to the underlying dynamics, provided 
such dynamics generates a slowly-varying mass function $M(N,Z)$. 
Moreover, we observe that {\sl any} smooth physical observable---not
only $M(N,Z)$---will satisfy the GK relations.

As noted in Eq.~(\ref{GKsApprox}), the GK relations  are proportional to
the {\sl crossed} (or mixed) derivatives of $M(N,Z)$. As any function of 
two variables has four independent third-order derivatives, two more 
``Garvey-Kelson'' relations may be derived to this order. That is, 
\begin{subequations}
 \begin{align}
 \Delta M_{4}^{(1)}(N,Z) &\equiv M(N+2,Z)-3M(N+1,Z)
 \nonumber \\ 
 & +3M(N,Z)-M(N-1,Z)
      \nonumber \\ & = \frac{\partial^{3}M}{\partial N^{3}}+
      \mathcal{O}(\partial^{4}M) \;, \label{GK3Approx}\\
 \Delta M_{4}^{(2)}(N,Z) &\equiv M(N,Z+2)-3M(N,Z+1)
 \nonumber \\ 
 & +3M(N,Z)-M(N,Z-1)
       \nonumber \\ & = \frac{\partial^{3}M}{\partial Z^{3}}+
       \mathcal{O}(\partial^{4}M) \;.  \label{GK4Approx}
 \end{align}
 \label{GK34Approx}
\end{subequations}
These last two relations correlate the masses of four nuclei along an
isotopic (constant $Z$) and isotonic (constant $N$) chain,
respectively.  Note, however, that the formalism is completely general
so one is free to select any two independent variables ({\sl e.g.,}
$A$ and $N\!-\!Z$) to carry out the analysis. Finally, note that the
approach may be systematically improved by selecting linear
combinations of masses so that third-order (and even
higher-order) derivatives get cancelled out. Such extended GK
relations afford higher accuracy in the prediction of nuclear masses,
albeit at the expense of having reliable information on the masses
of a larger number of neighboring nuclei.  Indeed, Barea and
collaborators have recently obtained a single relation involving the
masses of 21 nuclei; see Eq.~(4) of Ref.~\cite{Barea:2008zz}. If one
denotes such linear combination as $\Delta M_{21}(N,Z)$, one can show
that all derivatives up to fifth order get cancelled out! That is,
 \begin{equation}
 \Delta M_{21}(N,Z) = 
   2\frac{\partial^{6}M}{\partial Z^{2}\partial N^{4}}  
  +2\frac{\partial^{6}M}{\partial Z^{4}\partial N^{2}} 
  +\mathcal{O}(\partial^{7}M) \;.
   \label{GK21Approx}
 \end{equation}
Thus, one can systematically improve the accuracy of the original GK
relations by progressively removing third- and higher-order
derivatives.

\begin{table}[b]
\caption{Central value of the measured charge radii for 7 of 8
nuclei required to test the two Garvey-Kelson relations (\ref{GKs}) in
the region around ${}^{195}$Pb ($Z\!=\!82$, $N\!=\!113$), taken from
Ref.~\cite{Angeli:2004}. The charge radius of ${}^{195}$Pb is unknown.
Theoretical neutron radii obtained with the FSUGold
interaction~\cite{Todd-Rutel:2005fa} are also shown.}
\begin{ruledtabular}
\begin{tabular}{ccccccc}
Nucleus & $R_{\rm ch}$(fm) & $R_{n}$(fm) & &
Nucleus & $R_{\rm ch}$(fm) & $R_{n}$(fm)     \\
\hline
$^{197}_{~82}$Pb & 5.4420 & 5.5390 && $^{195}_{~80}$Hg & 5.4347 & 5.5355 \\
$^{193}_{~80}$Hg & 5.4239 & 5.5102 && $^{195}_{~82}$Pb & ------ & 5.5143 \\
$^{194}_{~80}$Hg & 5.4311 & 5.5219 && $^{194}_{~81}$Tl & 5.4233 & 5.5119 \\
$^{196}_{~81}$Tl & 5.4304 & 5.5369 && $^{194}_{~80}$Hg & 5.4311 & 5.5219 \\
$^{194}_{~81}$Tl & 5.4233 & 5.5119 && $^{196}_{~82}$Pb & 5.4420 & 5.5259 \\
$^{196}_{~82}$Pb & 5.4420 & 5.5259 && $^{196}_{~81}$Tl & 5.4304 & 5.5369 \\
\hline
$\vert\Delta R_{6}^{(2)}\vert$ & 0.0001 & 0.0002 &&
$\vert\Delta R_{6}^{(1)}\vert$ & 5.4385 & 0.0002 
\end{tabular}
\end{ruledtabular}
\label{Table1}
\end{table}

We next present two examples that aim to illustrate the generality and
flexibility of the approach. One of them goes beyond nuclear
masses and tests the validity of the GK relations for nuclear
radii. The other one is used to derive---or rather to
re-derive---some well-known mass relations among baryons. The chosen
examples are not intended to be exhaustive or systematic, but serve to
illustrate the potential of the GK relations.

\begin{figure}[b]
\vspace{-0.05in}
\includegraphics[height=2.50in,angle=0]{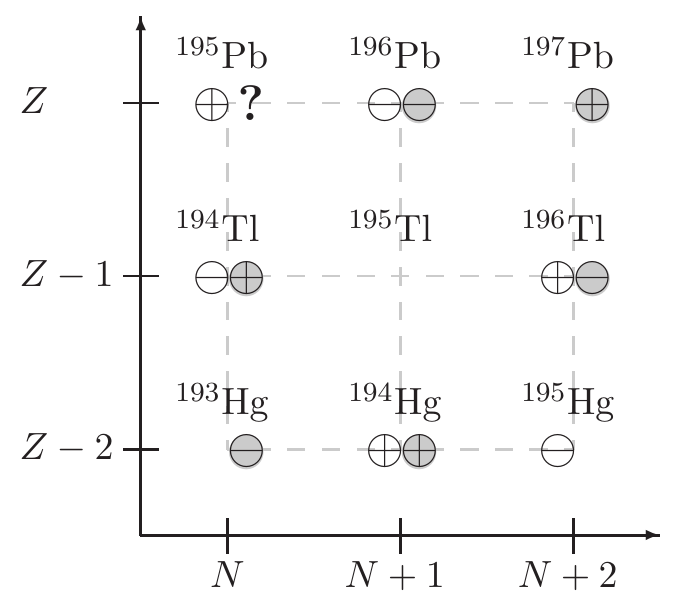}
\caption{Pictorial representation of the Garvey-Kelson relations in
the ${}^{195}$Pb region ($Z\!=\!82$, $N\!=\!113$). The unshaded
(shaded) circles represent six nuclei involved in
the first (second) Garvey-Kelson relation. The ``plus'' and ``minus''
signs inside the circles indicate how they should be included in the
GK relations. The nucleus ${}^{195}$Tl plays no role in this example.} 
\label{Fig1}
\end{figure}

Our first example concerns itself with nuclear radii. The charge
distribution of atomic nuclei can be accessed very cleanly with
electromagnetic probes and, indeed, the charge radii of many nuclei 
are known to exquisite accuracy. Although smaller than for masses, there
exists a sizable experimental database of nuclear charge radii (see
the 2004 compilation by Angeli~\cite{Angeli:2004}).  In applying the
GK relations to nuclear charge radii, we find a rms deviation of about
0.01~fm for the approximately 130 nuclei that satisfy (at least) one
of the two original GK relations [see Eq.~(\ref{GKs})].  This result
is consistent with the smooth global structure of the charge radius
identified in Ref.~\cite{Angeli:2004}. To provide a specific example,
we list in Table~\ref{Table1} the charge radii of 7 out of 8 nuclei
that are required to test the GK relations (\ref{GKs}) in the
$^{195}{\rm Pb}$ region. A pictorial representation is also provided
in Fig.~\ref{Fig1}. The available charge radii are sufficient to
verify that the GK relation (\ref{GK2}) is satisfied---at least in
this region---to a remarkably high accuracy ($\sim$10$^{-4}$~fm).
Having established the validity of one of the GK relations in this
region, one can then use the other relation to predict the central
value of the charge radius of ${}^{195}{\rm Pb}$ which is unknown
experimentally to be $R_{\rm ch}({}^{195}{\rm Pb})\!=\!5.4385 $~fm.
Note, however, that the GK relations (\ref{GKs}) applied to $^{195}$Pb
with different neighboring nuclei allow up to 12 independent
estimates~\cite{Barea:2008zz} for $R_{\rm ch}({}^{195}{\rm Pb})$. For
the present case, the experimentally available data limit these to
four. When all these four estimates are averaged, one obtains a value
of $R_{\rm ch}({}^{195}{\rm Pb})\!=\!5.437(5) $~fm. For completeness,
theoretically generated neutron radii in the same region have been
added to the table. Although unavailable experimentally, the close
fulfilment of the GK relations suggests that the underlying
neutron-radius function $R_{n}(N,Z)$ (whatever it may be) is likely to
be smooth.

In our second example we derive Garvey-Kelson relations for baryon
masses. For simplicity, we focus exclusively on the baryon octet and
decuplet. To apply the GK relation to baryons, it is sufficient to
replace the neutron and proton numbers in Eqs.~(\ref{GKs})
and~(\ref{GK34Approx}) with the strangeness ($N\rightarrow S$) and
electric charge ($Z\rightarrow Q$) of the baryon, respectively, and to
expand around $S\!=\!-2$ and $Q\!=\!1$. As there is no double-strange
baryon of positive charge, the GK relation (\ref{GK1}) is not
applicable. However, the relation (\ref{GK2}) may be applied and
yields the interesting result
\begin{align}
   \Delta_{\rm CG} &= 
   (p-n) - (\Sigma^{+}-\Sigma^{-}) + (\Xi^{0}-\Xi^{-})
   \nonumber \\
   &=\frac{\partial^{3}m_{8}}{\partial Q^{2}\partial S}
      +\frac{\partial^{3}m_{8}}{\partial Q\partial S^{2}}  
      +\mathcal{O}(\partial^{4}m_{8}) \;,
 \label{ColemanGlashow}
\end{align}
where $m_{8}(S,Q)$ is the underlying ground-state baryon octet mass
function. The equation $\Delta_{\rm CG} \!=\!0$ is the celebrated
Coleman-Glashow (CG) mass relation~\cite{Coleman:1961jn}; derived
originally using unbroken flavor SU(3)\cite{Coleman:1961jn} and later
also by methods such as the $1/N_c$
expansion~\cite{Jenkins:1995td,Capstick:2000qj,Jenkins:2000mi} and the
QCD parametrization method~\cite{Dillon:2000cu}. The CG mas relation
vanishes within experimental
error~\cite{Jenkins:1995td,Capstick:2000qj,Jenkins:2000mi,Dillon:2000cu}.
This suggests that the underlying mass function
$m_{8}(S,Q)$---although likely unknown---is smooth enough to justify
the use of the GK relations in the baryon sector.  However, to
this order, the GK relations yield no further information because the
structure of the octet is too limited for the application of
Eq.~(\ref{GK34Approx}). 

Although the first of the GK relations is still not applicable in the
case of the ($J^{\pi}=3/2^{+}$) baryon decuplet, the three other
relations (\ref{GK2}), (\ref{GK3Approx}), and~(\ref{GK4Approx})
yield interesting results, namely,
\begin{subequations}
 \begin{align}
  &\Delta^{*}_{\rm CG}  = (\Delta^{+}-\Delta^{0}) 
                          - (\Sigma^{*+}-\Sigma^{*-})
                          + (\Xi^{*0}-\Xi^{*-}) \approx 0 \;, \\
  &\Delta_{Q=-1} = \Delta^{-} -3\Sigma^{*-} + 3\Xi^{*-} - \Omega^{-}
                        \approx 0 \;, \\
  &\Delta_{S=0} = \Delta^{++} - 3\Delta^{+}  + 3\Delta^{0} - \Delta^{-}  
                       \approx 0 \;.
 \end{align}
 \label{GKDecuplet}
\end{subequations}
Note that the approximate equal sign implies that all three
expressions are proportional to a {\sl third derivative} of an
underlying (decuplet) mass function that is assumed to be slowly
varying. Indeed, all three equations vanish to within experimental
errors.  The first of the three equations is the corresponding
Coleman-Glashow relation extended to the baryon decuplet. The second
{\sl ``isocharge''} ($Q\!=\!-1$) relation connects the masses of all
negatively-charged baryons in the decuplet. Finally, the last {\sl
``isostrangeness''} ($S\!=\!0$) relation connects the masses of the
four $\Delta$ isobars in the multiplet (this relation is referred as
$\Delta_{3}$ in Ref.~\cite{Jenkins:1995td}). Note that there is an
additional relation connecting four of the baryons in the decuplet,
{\sl vis-\`a-vis},
\begin{equation}
 \Delta_{Q-S=2} = \Delta^{++} - 3\Sigma^{*+}  + 3\Xi^{*0} - \Omega^{-}
                        \approx 0 \;;
  \label{GKDecuplet2}
\end{equation} 
however, it is not independent of the other three.

It should be noted that essential to the derivation of the mass
relations is only the structure of the multiplets, namely octet and
decuplet, but not the particular use of the ground-state
baryons. Indeed, identical mass relations will hold true whether the
hadrons are baryons or mesons or whether they contain strange or
charmed quarks. Thus, we conclude this section with a remark on
the masses of charmed baryons. Replacing all strange quarks with charm
quarks in Eq.~(\ref{GKDecuplet2}) yields the following
($J^{\pi}=3/2^{+}$) mass relation:
\begin{equation}
 \Delta^{(c)}_{Q=2} = \Delta^{++} - 3\Sigma_{c}^{++}  
                           + 3\Xi_{cc}^{++} - \Omega_{ccc}^{++} 
                        \approx 0 \;.
  \label{GKCharm1}
\end{equation} 
Since at present neither the mass of the ($3/2^{+}$) double- and
triple-charm baryons are known, a prediction can be made only for a
specific linear combination. Using the most up to date information
from the Particle Data Group~\cite{Amsler:2008zzb} we obtain
\begin{equation}
  3\Xi_{cc}^{++} - \Omega_{ccc}^{++}  \approx
  3\Sigma_{c}^{++}  - \Delta^{++}  \approx 6328~{\rm MeV}\;.
  \label{GKCharm2}
\end{equation} 
Note that a recent theoretical study based on the Bethe-Salpeter equation 
yields a value of 6396 MeV for this combination \cite{Migura:2006ep},
or a 1\%  discrepancy.

In summary, although well motivated on physical grounds, we found
the Garvey-Kelson relations significantly more general and more
powerful than originally assumed. Using the simple---yet
little-known---fact that the linear combinations of masses that
enter into the GK relations are proportional to the third derivatives
of the nuclear mass function~\cite{deShalit:1974}, we
concluded that the validity of the GK relations hinges exclusively on
the smoothness of the underlying function and on nothing else. Having
established this fact, we regard the following three as the main 
conclusions of this work: (1) {\sl the GK relations are model
independent}; (2) {\sl any slowly-varying physical observable
satisfies the GK relations}; and (3) {\sl the accuracy of the GK
relations may be systematically improved}.

As examples chosen to illustrate the power and flexibility of the
approach, the GK relations were extended to the calculation of nuclear
radii and mass relations in the baryon sector. In the case of nuclear
radii, we focused on a region of the periodic table that enabled both
the validation of the approach and the prediction of the unknown
charge radius of ${}^{195}$Pb. For the case of baryon masses, a direct
application of the GK relations to the ground-state octet and decuplet
reproduced some well-known mass relations, such as the celebrated one
by Coleman and Glashow~\cite{Coleman:1961jn}. In addition, a
model-independent prediction was made for a linear combination of the
masses of the double- and triple-charm baryons that seems to be in
good agreement with available theoretical calculations.

 In the future we plan to a carry out a comprehensive and systematic
 program that will extend the Garvey-Kelson approach to areas that go
 beyond its original intent.
 However, we hope that the present study may already inspire the
 reader to consider applications in other fields.
We are confident that the approach can be
successfully generalized to the study of physical observables in
atomic, nuclear, and particle physics---and perhaps even to areas
outside of physics. The sole requirement for the Garvey-Kelson
relations to hold is a slowly-varying observable. Remarkably, nature
often seems to generate precisely such kind of observables independent
of the complicated---and likely unknown---underlying dynamics.

\begin{acknowledgments}
 J.P. is indebted to Profs. A. Frank and
 J. Hirsch for their hospitality during his sabbatical leave at the
 Instituto de Ciencias Nucleares (UNAM) and for enlightening
 discussions on the Garvey-Kelson relations. Work supported
 in part by grants from the U.S. DOE DE-FD05-92ER40750,
 FIS2008-01661 (Spain and FEDER), and 2005SGR-00343 (Spain),
 and by the Consolider-Ingenio 2010 Programme CPAN CSD2007-00042.
 J.P. and X.R. acknowledge grants 2008PIV00094 from AGAUR
 and AP2005-4751 from MEC (Spain), respectively.
\end{acknowledgments}

\bibliography{/Users/jorge/Tex/Papers/ReferencesJP}

\end{document}